\documentclass[12pt]{article}
\usepackage{amssymb}
\everymath={\displaystyle}

\newcommand{\be}{\begin{equation}}
\newcommand{\ee}{\end{equation}}

\date{December 2001}
\begin{document}
\begin{center}
{\large {\bf Darboux transformations for quasi-exactly
solvable Hamiltonians}}\\
\vspace{2cm}
 N. DEBERGH\footnote{email: Nathalie.Debergh@ulg.ac.be}(a),
Boris F. SAMSONOV\footnote{email: samsonov@phys.tsu.ru}(b) \\[1em] and
B. VAN DEN BOSSCHE\footnote{email: bvandenbossche@ulg.ac.be}(a)\\
\vspace{1cm}
(a){\it Fundamental Theoretical Physics,\\
Institute of Physics (B5),\\
University of Li\`ege,\\
B-4000 LIEGE  (Belgium)}\\ \vspace{0.5cm}

(b){\it Department of Quantum Field Theory,\\
Tomsk State University,\\
36 Lenin Ave., \\
634050 TOMSK (Russia)} \\
\end{center}
\vspace{0.5cm}
\begin{abstract}
We construct new quasi-exactly solvable one-dimensional potentials through
Darboux transformations. Three directions are investigated:
 Reducible and two types of irreducible second-order
 transformations. The irreducible transformations  of the first type give
 singular intermediate potentials and the ones of the second type
 give complex-valued intermediate potentials while final
 potentials are meaningful in all cases.
 These developments are
 illustrated on the so-called radial sextic oscillator.
\end{abstract}
%\newpage

\section{Introduction}
The resolution of the one-dimensional Schr\" odinger equation has
attracted much attention since the first statements of quantum mechanics.
It is now well known that this equation can be analytically solved for a
small number of interactions only. In that case, it is referred to as an
exactly solvable (ES) equation. Because of their major interest
(analytic solutions describe more finely the physical reality),
several methods have been proposed in order to increase the number of
these ES Schr\" odinger equations.
One can cite the factorization method introduced by Schr\" odinger
\cite{1} and actualized by Witten in the form of supersymmetric
quantum mechanics \cite{2}. It is based on the fact that a one-dimensional
Schr\" odinger Hamiltonian
\be
H_0 \equiv  -\frac{d^2}{dx^2}+V_0(x)
\label{eq1}
\ee
can be factorized as
\be
H_0=L_0^{\dagger}L_0 + \alpha\,,
\label{eq2}
\ee
with
\be
L_0=\frac{d}{dx} + W_0(x),
\label{eq3}
\ee
the constant $\alpha$ being called the factorization constant.
One can then obtain in a straightforward way the eigenfunctions of
\be
H_1 \equiv L_0 L_0^{\dagger} + \alpha =
-\frac{d^2}{dx^2}+V_1(x),
\label{eq4}
\ee
\be\label{eq3a}
V_1(x)=V_0(x)+2W_0'(x)
\ee
where the prime denotes the derivative with respect to $x$.
They are precisely given by
\be
\phi_N(x) = L_0 \psi_N(x) \, ,
\label{eq5}
\ee
where $\psi_N(x)$ stands for the eigenfunction (of eigenvalue $E_N$)
related to $H_0$.
The relation (\ref{eq4}) means that the operator $L_0^\dagger $
realizes the transformation in the opposite direction.
The function
\be
\widetilde\psi_N(x) = L_0^{\dagger} \phi_N(x) \, ,
\label{eq5a}
\ee
is an eigenfunction of $H_0$ differing from $\psi _N$ by a
normalization constant only. Note that neither $L_0$ nor $L_0^\dagger
$ are unitary and they do not preserve the norms of the functions.
The relations (\ref{eq5}) and (\ref{eq5a}) imply that $H_0$ and $H_1$ are
isospectral up to the fact that either the ground state of $H_0$
could belong to the kernel of the operator $L$ or the ground state of
$H_1$ could belong to the kernel of the operator $L^\dagger $.
In this case the spectra of $H_0$ and $H_1$ differ by the ground state
level only and they are strictly isospectral otherwise.
\par
Another method for giving rise to ES
Hamiltonians is the so-called Darboux transformation \cite{3}.
This method is based on a general notion of transformation
operators \cite{levitan}. The main element in definition
of such an operator is the
so-called intertwining relation
\be\label{intertw}
LH_0=H_1L\,.
\ee
The factorization method ({\it a fortiori} the supersymmetric quantum
mechanics) appeared long after the publication of the
Darboux paper which was unfamiliar to physicists. But since the
paper by Andrianov et. al. \cite{Andr} it was realized that the
methods are equivalent. The relation (\ref{intertw}) gives the
same expression  for the transformation operator as in the original
paper by Darboux \cite{3} which coincides with (\ref{eq3})-(\ref{eq5}) and
precises the function $W_0(x)$ to be the logarithmic derivative
\be
W_0(x)=-\frac{d \, ({\rm ln} \psi(x))}{dx}\,.
\label{eq6}
\ee
of a solution to the initial Schr\"odinger equation
\be
(H_0-\alpha)\psi(x)=0\,.
\label{eq8}
\ee
Note that $\alpha $ here is exactly the same factorization
constant as in Eqs. (\ref{eq2}) and (\ref{eq4}). The function
$\psi(x) $ is called the transformation function (sometimes the
factorization function).
The only condition imposed on $\psi(x) $ is its absence of zeros
inside the interval where the equation (\ref{eq8}) is solved.
This condition may be satisfied only if $\alpha \le E_0$
(see e.g. \cite{Sukumar}) where $E_0$ is the ground state energy of $H_0$ if
it has a discrete spectrum or the lower bound of the continuous spectrum
otherwise. For $\alpha=E_0$ the function $\psi(x) $ is evidently
the ground state function of $H_0$ and for $\alpha <E_0$ it is
an
unphysical solution of (\ref{eq8}).

It is evident that the Hamiltonian $H_1$ can be taken as the
initial Hamiltonian for the next transformation step etc. In
such a way one gets a chain of so called {\it reducible}
transformations. The main feature of such a chain is that every
element of the chain has a well-defined (this means essentially
self-adjoint) Hamiltonian acting in the same Hilbert space and
the resulting action of the whole chain may be obtained with the
 help of a higher order transformation operator only acting
 on solutions of the initial equation. A natural
question then arises. Is it possible to get
something new with respect to a chain of
transformations by defining an $n$th order transformation operator
as an $n$th order differential operator satisfying
the intertwining relation (\ref{intertw})? The answer is positive:
 there are the
{\it irreducible} chains of transformations.
 The main feature of
such chains is that despite of the fact that they  can also decompose
into a superposition of first order operators, some of the
corresponding intermediate Hamiltonians can not be defined
properly: They are either singular or non Hermitian. More precisely,
it was first established that some intermediate
potentials may have poles  inside the interval where the
initial Schr\"odinger equation is solved but the final potential
is regular \cite{Krein} (the supersymmetric interpretation
can be found in \cite{MPLA} and a generalization of Krein's statement
is in \cite{PLA}). In this paper we refer to this case as
irreducible transformations of the first type. Another
possibility to get an irreducible chain consists in choosing
complex factorization energies $\alpha _1=\alpha $ and
$\alpha _2=\overline\alpha $ (the bar over a symbol
means the complex conjugation). In this case some intermediate potentials
are complex but the final potential is real \cite{compl}.
We refer to this case as irreducible transformations of the second type.

Irreducible Darboux transformations have already
been exploited in order to give rise to a large amount of new ES
Schr\" odinger equations \cite{4a}, \cite{MPLA}-\cite{SB}.
However a new kind of Schr\" odinger equations appeared in the literature
some twenty years ago \cite{5}. These are the so-called quasi-exactly
solvable (QES) Schr\" odinger equations, i.e., those for which a finite
number of solutions can be analytically determined. These equations are
rather exceptional. A previous attempt to list them in an exhaustive way
has been performed by Turbiner \cite{6} on the basis of the link between
these equations and the $sl(2,R)$ vector fields preserving
a finite-dimensional space. However it is now understood that some
of these equations \cite{7} do not have any connection with the Lie
algebra $sl(2,R)$. Therefore it is interesting to find
other approaches giving rise to new QES equations.
Supersymmetric quantum mechanics is one of these approaches but
it has been mainly used in the context of
ES equations (for a recent review, see e.g. \cite{CKS}).
In what concerns QES equations there are only several papers on this subject
(see e.g. \cite{8}).
The purpose of this paper is to fill in this gap.
   Thus, in Sections 2 and 3, we focus
on the reducible and irreducible second-order Darboux transformations,
respectively. In each of these Sections we produce new QES
potentials starting from the same QES potential. For simplicity we
choose it as the one corresponding to the radial
sextic oscillator \cite{9} even if our
developments can evidently be applied to any of the already known QES
interactions.

\section{Reducible second-order Darboux transformations}
\label{section2}
Let us start with the one-dimensional Schr\" odinger Hamiltonian $H_0$
as given in Eq. (\ref{eq1}) with
\be
V_0(x)=a^2x^6-2a(2M+2s+1)x^2+\frac{4(s-1/4)(s-3/4)}{x^2}\,, \quad x
\in {\mathbb R}_0^+
\label{eq9}
\ee
which describes the radial sextic oscillator, a prototype
of QES systems. In Eq. (\ref{eq9}), $a$ and $s$ are positive constants while
the positive integer $M$ is related to the number (equal to $M+1$) of
 analytic
eigenfunctions. These functions are expressed as \cite{9}
\be
\psi_N(x)=\exp{(-\frac{a}{4}x^4)}x^{2s-\frac{1}{2}}P_M^{(N)}(x^2)\,,\quad
N=0,1,\ldots ,M
\label{eq10}
\ee
where
\be\label{eq10a}
P_M^{(N)}(x^2)=\sum_{n=0}^M
c_n^{(N)}x^{2n}\,.
\ee
The coefficients $c_n^{(N)}$ have to be determined through the
resolution of a system of algebraic equations. Precisely these
coefficients are
\begin{eqnarray}
c_n^{(N)} &=& \frac{(-1)^{M-n}}{(M-n)!} Y_{M-n}^{(N)} (\alpha)\\
&\equiv& 1 \; {\rm if} \; n=M \nonumber \\
&\equiv& \frac{(-1)^{M-n}}{(M-n)!}\sum_{i_1\neq \ldots \neq i_{M-n}}
\alpha_{i_1}^{(N)}\ldots \alpha_{i_{M-n}}^{(N)} \; {\rm if}\; n \neq M,
\nonumber
\label{eq11}
\end{eqnarray}
with the $\alpha_{j}^{(N)}$'s being known through \cite{9}
\be
\sum_{k=1}^M \frac{(4ax^4-8s)}{x^2-\alpha_{k}^{(N)}}-\sum_{k\neq l =1}^M
\frac{4x^2}{(x^2-\alpha_{k}^{(N)})(x^2-\alpha_{l}^{(N)})}-4aMx^2-E_N=0.
\label{eq12}
\ee
\par
Let us now turn to the Darboux transformation. In the reducible case
we are concerned with, it is a product of two Darboux transformations as
given in Eq. (\ref{eq3}) with $W_0(x)$
being fixed according to Eq. (\ref{eq6}).
We first choose $\psi(x)$ to be the ground state of $H_0$ in order to prove
that even in this simplified context we can generate new QES potentials.
Indeed with such a choice we obtain
\be
W_0(x)=ax^3-\frac{(2s-1/2)}{x}-\sum_{k=1}^M\frac{2x}{(x^2-
\alpha_{k}^{(0)})}\,.
\label{eq13}
\ee
Therefore one can immediately obtain $M$ eigenfunctions
$\phi_N(x)$ (as expressed in Eq. (\ref{eq5})) of
the Hamiltonian $H_1$ defined in Eqs. (\ref{eq4}) and
(\ref{eq3a})
\be
\phi_N(x)=\frac{\exp{(-\frac{a}{4}x^4)}x^{2s-
\frac{1}{2}}}{P_M^{(0)}(x^2)} W(P_M^{(0)}(x^2),P_M^{(N)}(x^2)), \quad
N=1,2,\ldots ,M.
\label{eq15}
\ee
The symbol $W$ stands for the usual Wronskian i.e.
\be
W(f_1(x),f_2(x)) \equiv f_1(x)\frac{d f_2 (x)}{dx}-
f_2(x)\frac{d f_1(x)}{dx}\,.
\label{eq16}
\ee
Taking $f_1(x)$ and $f_2(x)$ being respectively
the polynomials $P_M^{K}(x^2)$ and
$P_M^{N}(x^2)$ this Wronskian will be denoted by $W_{KN}(x)$.
It is also clear from (\ref{eq15})
that $\phi_0(x)=0$ in accordance with Eq. (\ref{eq5}).
\par
The potential $V_1(x)$ related to these $M$ nontrivial solutions
$\phi_N(x)$ is expressed through Eq. (\ref{eq3a}) as
\begin{eqnarray}
V_1(x)&=&a^2x^6-4a(M+s-1)x^2+\frac{(4s^2-1/4)}{x^2}+
4\sum_{k=1}^M\frac{(2s+M-a(\alpha_k^{(0)})^2)}{x^2-\alpha_k^{(0)}}
\nonumber \\
&+&8\sum_{k=1}^M\frac{\alpha_k^{(0)}}{(x^2-\alpha_k^{(0)})^2}+
2\sum_{k\neq l=1}^M\frac{\alpha_k^{(0)}+
\alpha_l^{(0)}}{(x^2-\alpha_k^{(0)})(x^2-\alpha_l^{(0)})}\,.
\label{eq17}
\end{eqnarray}
It is thus a QES potential and being not subtended by $sl(2,R)$,
it has not been listed in \cite{6}. Yet at this stage we have thus
produced a new QES potential. This is a generalization of a
previously obtained potential \cite{8} which corresponds to the
choice
$a=1/2$ and $s=1/4$.
\par
We now proceed further by going to the second first-order Darboux operator
$L_1$ such that
\be
H_1=-\frac{d^2}{dx^2}+V_1(x)=L_1^{\dagger}L_1+E_1\,,
\label{eq18}
\ee
with
\be
L_1=\frac{d}{dx}+W_1(x)
\equiv \frac{d}{dx}-\frac{d ( {\rm ln}\phi_1(x))}{dx}\,.
\label{eq19}
\ee
We will then obtain once again a new QES potential $V_2(x)$ defined
according to Eq.(\ref{eq3a})
\be
V_2(x)=V_1(x)+2W_1^\prime (x)
\label{eq20}
\ee
and the corresponding $(M-1)$ eigenfunctions
\be
\chi_N(x) \equiv L_1 \phi_N(x) \, ,\quad N=2,...,M.
\label{eq21}
\ee
The function $W_1(x)$ characterizing this second Darboux operator is
determined according to Eqs. (\ref{eq19}) and (\ref{eq15}) and it is given
by
\be
W_1(x)=-W_0(x)+(E_1-E_0) \frac{P_M^{(0)}(x^2)P_M^{(1)}(x^2)}{W_{01}(x)}.
\label{eq22}
\ee
It is then straightforward to obtain $V_2(x)$ and $\chi_N(x)$ through
Eqs. (\ref{eq20}) and (\ref{eq21}), respectively. They are
\be\label{eq23}
V_2(x)=V_0(x)+
2(E_1-E_0)\frac{d}{dx}(\frac{P_M^{(0)}(x^2)P_M^{(1)}(x^2)}{W_{01}(x)})
\ee
and
\begin{eqnarray}
\chi_N(x)&=&\exp{(-\frac{a}{4}x^4)}
\frac{x^{2s-1/2}}{W_{01}(x)}\hfill \\
&&\times [-E_0P_M^{(0)}(x^2)W_{1N}(x)+
E_1P_M^{(1)}(x^2)W_{0N}(x)-E_NP_M^{(N)}(x^2)W_{01}(x)]\, ,\nonumber \\
&& N=2,\ldots ,M\,.\nonumber
\label{eq24}
\end{eqnarray}
%
%\begin{eqnarray}
%&&\chi_N(x)=\exp{(-\frac{a}{4}x^4)}
%\frac{x^{2s-1/2}}{W_{01}(x)}[-E_0P_M^{(0)}(x^2)W_{1N}(x)\nonumber \\
%&&+E_1P_M^{(1)}(x^2)W_{0N}(x)-E_NP_M^{(N)}(x^2)W_{01}(x)]\, ,
%\quad N=2,\ldots ,M\,.
%\label{eq24}
%\end{eqnarray}
%
Once again it is immediate to observe from Eq. (\ref{eq24})
that $\chi_0(x)=\chi_1(x)=0$. We thus obtain in Eq. (\ref{eq24})
the $(M-1)$ eigenfunctions of the new QES potential $V_2(x)$ given
in Eq. (\ref{eq23}).
\par
As stated in the Introduction, this succession of two first-order Darboux
operators is equivalent to a unique second-order Darboux operator defined by
\be
L \equiv L_1L_0\,.
\label{eq25}
\ee
Following the general theory of Darboux transformations \cite{3}, and this
can be directly checked through Eqs.
(\ref{eq3}), (\ref{eq13}), (\ref{eq19}) and (\ref{eq22}), this operator
is such that
\be
L \psi_N(x)=W^{-1}(\psi_0(x), \psi_1(x))W(\psi_0(x), \psi_1(x), \psi_N(x))
\label{eq26}
\ee
giving again the eigenfunctions $\chi_N(x)$ defined in Eq. (\ref{eq24}),
 while
\be
V_2(x)=V_0(x)-2\frac{d^2}{dx^2}\,[\,{\rm ln} W(\psi_0(x),\psi_1(x))\,]\,.
\label{eq27}
\ee
This Darboux operator $L$ is thus the one connecting the two QES
potentials $V_0(x)$ and $V_2(x)$ given in (\ref{eq23}). The
intermediate step $V_1(x)$ (see formula (\ref{eq17})) is also physically
meaningful. It is precisely the statement of reducible Darboux
transformations but adapted here to QES equations.

We observe here the usual situation inherent for the potentials
with the centrifugal term of the type $l(l+1)/x^2$, ($l=2s-3/2$). The
Darboux transformation results in the change $l \rightarrow l+1$
for this term even if it is not present in the initial
potential (i.e. $l=0$). This is due to zero boundary condition
for the functions (\ref{eq10}) and their specific asymptotic
behaviour when $x\to 0 $, $\psi_N(x)\sim x^{l+1}$.
\par
We end this Section by a specific example.
This will be in particular useful in order to compare the results we
have obtained here with the ones of the irreducible context in the next
Section. We thus take
\be
a=\frac{1}{2}\,,\quad  M=2
\label{eq28}
\ee
so that the starting potential $V_0(x)$ is
\be
V_0(x)=\frac{1}{4}x^6-(5+2s)x^2+\frac{4(s-1/4)(s-3/4)}{x^2}\,.
\label{eq29}
\ee
The corresponding eigenfunctions and eigenvalues are given by
\begin{eqnarray}
\psi_0(x)=&\exp{(-\frac{1}{8}x^4)}x^{2s-1/2}(x^4+2\sqrt{4s+1}x^2+4s) \,
, & E_0=-4\sqrt{4s+1}\,,
\label{eq30}\\
\psi_1(x)=&\exp{(-\frac{1}{8}x^4)}x^{2s-1/2}(x^4-4s-2) \,,
\hphantom{\sqrt{4s+1}x^2} &  E_1=0\,,
\label{eq31}\\
\psi_2(x)=&\exp{(-\frac{1}{8}x^4)}x^{2s-1/2}(x^4-2\sqrt{4s+1}x^2+4s) \, ,
& E_2=4\sqrt{4s+1}
\label{eq32}
\end{eqnarray}
as seen from Eq. (\ref{eq10}). The new QES potential $V_2(x)$ resulting from
the second-order reducible operator $L$ given by (\ref{eq25})
is (see Eq. (\ref{eq23}))
\begin{eqnarray}
V_2(x)&=&\frac{1}{4}x^6-(2s-1)x^2+\frac{8s}{x^2}+\frac{8(x^2-
\sqrt{4s+1})}{(x^4+2\sqrt{4s+1}x^2+4s+2)}\nonumber \\
&-&\frac{32x^2}{(x^4+2\sqrt{4s+1}x^2+4s+2)^2},
\label{eq33}
\end{eqnarray}
while its unique analytic eigenfunction $\chi_2(x)$ as well as
the corresponding energy $E_2$ are
\be
\chi_2(x)=\exp{(-\frac{1}{8}x^4)}
\frac{x^{2s+3/2}}{(x^4+2\sqrt{4s+1}x^2+4s+2)} \,, \quad E_2=4\sqrt{4s+1}\,.
\label{eq34}
\ee
Moreover there exists in this case a meaningful intermediate step
characterized by the potential $V_1(x)$, i.e.,
\begin{eqnarray}
V_1(x)&=&\frac{1}{4}x^6-2(s+1)x^2+
\frac{(4s^2-1/4)}{x^2}+\frac{4(1+\sqrt{4s+1})}{x^2+\sqrt{4s+1}-1}\nonumber
\\
&+&\frac{4(1-\sqrt{4s+1})}{x^2+\sqrt{4s+1}+1}
+\frac{8(1-\sqrt{4s+1})}{(x^2+\sqrt{4s+1}-1)^2}\nonumber \\
&-&\frac{8(1+\sqrt{4s+1})}{(x^2+\sqrt{4s+1}+1)^2}-
\frac{8\sqrt{4s+1}}{x^4+2\sqrt{4s+1}x^2+4s}.
\label{eq35}
\end{eqnarray}
Two eigenfunctions and eigenvalue related to this potential are also known.
They are given by
\begin{eqnarray}
\phi_1(x)=&\exp{(-\frac{1}{8}x^4)}x^{2s+1/2}
\frac{x^4+2\sqrt{4s+1}x^2+4s+2}{x^4+2\sqrt{4s+1}x^2+4s} \, ,
& E_1=0\,,
\label{eq36}\\
\phi_2(x)=&\exp{(-\frac{1}{8}x^4)}x^{2s+1/2}
\frac{x^4-4s}{x^4+2\sqrt{4s+1}x^2+4s} \, ,\hphantom{mm}
& E_2=4\sqrt{4s+1}\,.
\label{eq36a}
\end{eqnarray}
\section{Irreducible second-order Darboux\protect\newline transformations}
\label{section3}
\subsection{Transformations of the first type}

We now turn to the irreducible context.
We still consider at the start the potential $V_0(x)$ given in
Eq. (\ref{eq9}) as well as the corresponding eigenfunctions $\psi_N(x)$,
$N=0,1,\ldots ,M$ specified in Eq. (\ref{eq10}). The main purpose here is
to find an operator similar to the one given in Eq. (\ref{eq26}) but being
such that the factorization (\ref{eq25}) cannot give rise to a physically
meaningful ${\rm V_1}(x)$. We thus propose to consider an operator
${\rm L}$ being built on $\psi_1(x)$ and $\psi_2(x)$ i.e.
\be
{\rm L} \psi_N(x)\equiv W^{-1}(\psi_1(x), \psi_2(x))
W(\psi_1(x), \psi_2(x), \psi_N(x))
\label{eq38}
\ee
such that ${\rm L}\psi_N(x)$ will stand for the $(M-1)$
($N=0,3,4,\ldots ,M$) eigenfunctions of the new QES potential replacing
$V_2(x)$ given in (\ref{eq27}).
\par
One can see, comparing Eq. (\ref{eq38}) with Eq. (\ref{eq26}), that the
irreducible case can be deduced from the reducible one by the simple
changes of indices: $0 \rightarrow 1$ and $1 \rightarrow 2$,
in Eqs. (\ref{eq26}) and (\ref{eq27}) respectively.
Through example (\ref{eq33}) we thus  get
\begin{eqnarray}
{\rm V_2}(x)&=&\frac{1}{4}x^6-(2s-1)x^2+\frac{8s}{x^2}+
\frac{8(x^2+\sqrt{4s+1})}{(x^4-2\sqrt{4s+1}x^2+4s+2)}\nonumber \\
&-&\frac{32x^2}{(x^4-2\sqrt{4s+1}x^2+4s+2)^2}
\label{eq39}
\end{eqnarray}
and
\be
\chi_0(x)=\exp{(-\frac{1}{8}x^4)}
\frac{x^{2s+3/2}}{(x^4-2\sqrt{4s+1}x^2+4s+2)} \, , \quad
E_0=-4\sqrt{4s+1}\,.
\label{eq40}
\ee
One immediately notice that in this irreducible case,
the potential (\ref{eq39}) as well as its eigenfunction (\ref{eq40}) can be
deduced from their reducible analogues (\ref{eq33}) and (\ref{eq34}) by a
simple change of sign in front of the square roots.
However it is a new QES potential with a unique eigenfunction $\chi_0(x)$
given in Eq. (\ref{eq40}) and characterized this time by the lowest energy
of the potential (\ref{eq9}) when restricted to the choices (\ref{eq28}).
\par
As stated before, the Darboux operator ${\rm L}$ cannot give rise to a
physically relevant intermediate potential ${\rm V_1}(x)$. Indeed,
the operator ${\rm L}$ defined in Eq. (\ref{eq38}) can be written as
${\rm L}= {\rm L_1}{\rm L_0}$
%
%\be
%{\rm L}= {\rm L_1}{\rm L_0}
%\label{eq41}
%\ee
%
with
\be
{\rm L_j} \equiv \frac{d}{dx}+{\rm w_j}(x),\quad j=0,1\,.
\label{eq42}
\ee
More precisely, the function ${\rm w_0}(x)$, the irreducible analogue of
$W_0(x)$ given in Eq. (\ref{eq13}), is determined as
\be
{\rm w_0}(x)=\frac{1}{2}x^3-\frac{(2s-1/2)}{x}-
2x\left(\frac{1}{x^2-\sqrt{4s+1}-1}+\frac{1}{x^2-\sqrt{4s+1}+1}\right),
\label{eq43}
\ee
which effectively coincides with the one of Eq. (\ref{eq13}) up to the
choices (\ref{eq28}) and the replacement of $\alpha_k^{(0)}$
($k=1,\ldots ,M$) by $\alpha_k^{(2)}$. As clear from Eq. (\ref{eq43}),
this leads to singularities which were not present in the reducible case
(\ref{eq13}). These singularities are also recovered at the step of the
potential ${\rm V_1}(x)$ obtained through Eq. (\ref{eq3a}).
%
%\be
%{\rm V_1}(x) \equiv {\rm w_0}^2(x) + \frac{d{\rm w_0}(x)}{dx}+E_2\,.
%\label{eq44}
%\ee
%
Indeed we have (compare with Eq. (\ref{eq35}))
\begin{eqnarray}
{\rm V_1}(x)&=&\frac{1}{4}x^6-2(s+1)x^2+\frac{(4s^2-1/
4)}{x^2}+\frac{4(1+\sqrt{4s+1})}{x^2-\sqrt{4s+1}+1}\nonumber \\
&+&\frac{4(1-\sqrt{4s+1})}{x^2-\sqrt{4s+1}-1}
+\frac{8(1+\sqrt{4s+1})}{(x^2-\sqrt{4s+1}-1)^2}\nonumber \\
&-&\frac{8(1-\sqrt{4s+1})}{(x^2-\sqrt{4s+1}+1)^2}+
\frac{8\sqrt{4s+1}}{x^4-2\sqrt{4s+1}x^2+4s}.
\label{eq45}
\end{eqnarray}
This potential is ill-defined in the space ${\mathbb R}_0^+$
we are working in and cannot be considered as a convenient intermediate step
proving then the irreducible character of ${\rm L}$.

\subsection{Transformations of the second type}
\label{section4}
Let us now conclude with the third specific feature of the Darboux
transformation namely the fact that one can construct physically
significant new potentials from complex solutions of the initial equation.
Once again we concentrate on the radial sextic oscillator characterized
by the potential $V_0(x)$ given in Eq. (\ref{eq9}). In order to illustrate
our approach, we only need to find non-physical eigenfunctions
corresponding to complex eigenvalues.

We shall now show  that the potential (\ref{eq9}) with the integer and
half-integer values of $s$ has analytic solutions for some complex values of
the "energy" (for simplicity we shall continue to use
this terminology). For this purpose we note that this equation is
covariant under the transformation $s\to 1-s$, $M\to M+2s-1$.
This means that for a given
value of the positive integer $M$ and some values of the
integer or half-integer $s$, the
eigenfunction of $H_0$ with $s$ replaced by $\widetilde s=1-s$ and
$M$ by $\widetilde M=M+2s-1$ is also the eigenfunction of $H_0$ with
the given $M$ and $s$. Note that $\widetilde s$ is negative for
$s>1$.
The values of the energy calculated according to \cite{9} with negative $s$
 might be either complex or real. When they are real they lead
 to known solutions but for complex values of the energy we
 obtain new solutions of the Schr\"odinger equation. These new
 solutions when used to construct second order transformations lead
 to new real QES potentials.
The second-order
Darboux operator $\Lambda$ is defined in complete analogy
with Eq. (\ref{eq26}) by
\be
\Lambda \psi_0(x)= W^{-1}(\psi(x),\bar{\psi}(x))
W(\psi(x),\bar{\psi}(x),\psi_0(x))\,.
\label{eq52}
\ee
The state (\ref{eq52}) will then be the eigenfunction
(of the same eigenvalue as the one of $\psi _0(x)$ ) of
the QES potential
\be
\nu_2(x)=V_0(x)-2\frac{d^2}{dx^2}({\rm ln} W(\psi(x),\bar{\psi}(x)))\,.
\label{eq53}
\ee

 Let us illustrate now this possibility by a specific example.
We take $a=1/2$, $M=0$ and $s=2$.
In this case
$\widetilde M=3$ and $\widetilde s=-1$. The complex eigenfunction
of the potential
\be\label{le}
V_0(x)=\frac{1}{4}x^6-5x^2+\frac{35}{4x^2}
\ee
 is
$$
\psi (x)=x^{-5/2}\exp
(-x^4/8)[\frac{1}{2}x^2(\frac{1}{2}x^4-5)+i\sqrt{5}(\frac{1}{2}x^4-1)]\,.
$$
It corresponds to the energy $E=-i4\sqrt{5}$. The
single analytic eigenfunction (corresponding to $E_0=0$)
of this potential (\ref{le}) is
\be
\psi_0(x)=\exp{(-\frac{1}{8}x^4)}x^{\frac{7}{2}}\,.
\label{eq48}
\ee

The new QES potential reads
\be\label{Vtransf2}
\nu_2(x)=\frac{1}{4}x^6+x^2+\frac{3}{4x^2}+16\frac{-6x^2+x^6}{20+4x^4+x^8}-
2048\frac{x^6}{(20+4x^4+x^8)^2}\,
\ee
and its solution is
\be \Lambda \psi_0(x)=
\frac{\exp{(-\frac{1}{8}x^4)}x^{\frac{3}{2}}(x^4+6)}
{x^8+4x^4+20} \, , \quad E_0=0 \,.\label{eq54}
\ee

\vspace{.5em}
\noindent
{\bf\Large{Acknowledgments}}

\vspace{.5em}
\noindent
The work of BFS is partially supported by the Russian Foundation
for Basic Research.
BFS is also grateful to Belgium Fonds National de la Recherche
Scientifique for a financial support during his visit at
Liege University in Autumn 2001. The work of ND and BVdB was
supported by l'Institut Interuniversitaire des Sciences
Nucl\' eaires (Belgium).

\end{document}